# Tunable Mie Resonances in the Visible Spectrum

Li Lu[1,#], Zhaogang Dong[2,#,*], Febiana Tijiptoharsono[2], Ray Jia Hong Ng[3], Hongtao Wang[1], Soroosh Daqiqeh Rezaei[1], Yunzheng Wang[1], Hai Sheng Leong[2], Joel K. W. Yang[1,2,*] and Robert E. Simpson[1,*]

[1]Singapore University of Technology and Design, 8 Somapah Road, 487372, Singapore

[2]Institute of Materials Research and Engineering, A*STAR (Agency for Science, Technology and Research), 2 Fusionopolis Way, #08-03 Innovis, 138634 Singapore

[3]Institute of High Performance Computing, A*STAR (Agency for Science, Technology and Research), 1 Fusionopolis Way, #16-16 Connexis, 138632 Singapore

[#]These authors equally contribute to this work.

*Correspondence and requests for materials should be addressed to R.E.S. (email: robert_simpson@sutd.edu.sg), J.K.W.Y. (email: joel_yang@sutd.edu.sg) and Z.D. (email: dongz@imre.a-star.edu.sg).

**ABSTRACT**

Dielectric optical nanoantennas play an important role in color displays, metasurface holograms, and wavefront shaping applications. They usually exploit Mie resonances as supported on nanostructures with high refractive index, such as Si and $TiO_2$. However, these resonances normally cannot be tuned. Although phase change materials, such as the germanium-antimony-tellurium alloys and post-transition metal oxides, such as ITO, have been used to tune optical antennas in the near-infrared spectrum, tunable dielectric antennae in the visible spectrum remain to be demonstrated. In this paper, we designed and experimentally demonstrated tunable dielectric nanoantenna arrays with Mie resonances in the visible spectrum, exploiting phase transitions in wide-bandgap $Sb_2S_3$ nano-resonators. In the amorphous state, Mie resonances in these $Sb_2S_3$ nanostructures give rise to a strong structural color in reflection mode. Thermal annealing induced crystallization and laser induced amorphization of the $Sb_2S_3$ resonators allow the color to be tuned reversibly. We believe these tunable $Sb_2S_3$ nanoantennae arrays will enable a wide variety of tunable nanophotonic applications, such as high-resolution color displays, holographic displays, and miniature LiDAR systems.

**Introduction**

Dielectric optical nanoantennas interact strongly with light through both electric dipole and magnetic dipole resonances.[1-3] Directional scattering arises when these resonances are interacting with each other either constructively or destructively, known as the 1$^{st}$ or 2$^{nd}$ Kerker's conditions.[4, 5] Dielectric optical nanoantennas have been used to demonstrate holograms,[6-8] color nanoprints,[3, 9-11] miniaturized lasers,[12, 13] localized field enhancements,[14] hybrid metal-dielectric antenna,[15] high-quality factor resonators,[16, 17] and optical beam steering.[18, 19] However, all of these demonstrations were non-tunable because high refractive index materials, such as Si, TiO$_2$ and GaP,[20] are typically used and the optical constants of these high index dielectric materials cannot be actively tuned.

There are two ways to achieve dynamic control of Mie resonance on dielectric nanoantennas. One way is to tune the resonances by controlling the local environment of Mie resonators, such as using liquid crystals (LCs).[21, 22] For instance, Si dielectric nanoantennas with a bandgap of 1.1 eV can be embedded in a liquid crystal. When the liquid crystal is switched from nematic to isotropic phase, its resonances can be switched over a narrow band around 1550 nm (*i.e.*, 0.8 eV).[21, 22] In addition, the liquid crystal setup is volatile and it requires a constant energy to keep its state. Another way is to tune the Mie resonator itself by using materials with tunable optical properties, such as vanadium dioxide (VO$_2$), where its metal-insulator phase transition has been used to demonstrate electrically-tuned beam steering at the RF regime of ~100 GHz.[23] However, these approaches are limited to the infrared or radio frequency (RF) regime. Therefore, there have been no practical demonstrations of tunable Mie resonances in the visible spectrum until now and tuning resonances in visible spectrum has remained elusive.

Integrating phase change materials (PCMs) into dielectric nanoantennas provides a non-volatile and energy-efficient approach to achieve tunable optical devices.[24-27] This non-volatile characteristic originates from the high sensitivity of the optical constants to atomic reconfigurations.[27] Among all the PCMs, $Ge_2Sb_2Te_5$ (GST)[24, 28, 29] and $Ge_2Sb_2Se_4Te$ (GSST)[30] are commonly investigated because of their success in phase change optical memory and data storage.[31] However, these data storage PCMs possess small bandgaps of ~0.5 eV[32] and consequently absorb visible light, reducing the quality-factor of the resonances. Therefore, resonance tuning demonstrations have been limited to the infrared spectrum. Recently, we proposed antimony trisulfide ($Sb_2S_3$) as an alternative wide bandgap PCM for phase change programmable photonics in the visible spectrum.[33] $Sb_2S_3$ has a bandgap of 1.70 to 2.05 eV, moving the absorptance band-edge to the middle of the visible spectrum at ~600 nm. More importantly, the refractive index of $Sb_2S_3$ is high, and it can be tuned from 3 to 3.5 in the visible spectrum.[33] Until now, Fabry-Perot-like resonance have only been studied in multi-layer $Sb_2S_3$ thin film stacks.[33-35] We hypothesized, therefore, that this characteristic of low absorption, high refractive index, and refractive index tunability in the visible spectrum makes $Sb_2S_3$ ideal for programming Mie resonances. Recently, similar hypothesis and theoretical studies have been reported, but a practical device with reversible switching have yet to be demonstrated.[36]

In this paper, we describe the design, fabrication, and characterization of a tunable $Sb_2S_3$ dielectric nanoantenna array with Mie resonances in the visible spectrum. In the amorphous state, the nanostructured $Sb_2S_3$ antennae exhibit strong reflected structural color. The Mie resonances and concomitant reflected colors can be reversibly switched using hotplate thermal annealing to crystallize and laser heat pulses to amorphize the nanoantennas. These demonstrable manipulations of Mie resonances in the visible spectrum are the first step toward high-resolution color displays,[37] holographic displays[38] and miniature LiDAR scanning systems.[18, 19]

**Results**

We designed the all-dielectric $Sb_2S_3$ Mie resonator arrays based on the developed process to fabricate $Sb_2S_3$ nanodiscs with radii as small as 25 nm, where such small nanostructures resonate in the visible spectrum due to the high refractive index of $Sb_2S_3$. For instance, Figure 1(a) shows the detailed resonator schematic, where $Sb_2S_3$ nanostructures were fabricated on top of 200-nm-thick $SiO_2$ on silicon substrate. To fabricate the $Sb_2S_3$ nanodiscs, a 20-nm-thick palladium (Pd) layer was first patterned and used as a hard mask for the dry etching process. Reactive ion etching with $Cl_2$ gas chemistry[17] was then used to remove the exposed $Sb_2S_3$ (see details in the Methods section). Crucially, a 70-nm-thick $Si_3N_4$ film was then deposited onto the surface of resonator array to prevent the loss of the volatile sulfur (S) element during the phase transition process. At the same time, this 70-nm-thick $Si_3N_4$ film also acts as an anti-reflection coating for the exposed substrate. Embedding the structures within a dielectric media allows the $Sb_2S_3$ nanostructures to mimic its behavior in a "free-space" environment and leads to sharper spectral transitions at the points that satisfy Kerker's conditions.[3] The anti-reflection coating design utilizes the 2nd order anti-reflection resonance, which is slightly different from the 1st order anti-reflection mode (see the detailed explanation in Fig. S1).[3]

The optical constants of $Sb_2S_3$ in both the amorphous and crystalline states are shown in Figure 1(c)-(d). The refractive index contrast between the amorphous and crystalline state, $\Delta n$, is ~0.5 in the visible and near-infrared spectrum. Comparing with LCs, the refractive index change is usually less than 0.4.[21, 22] Figure 1(d) shows that amorphous $Sb_2S_3$ is essentially transparent for the wavelength longer than 600 nm. When $Sb_2S_3$ is switched into the crystalline state, $n$ increases, and it has an optical absorption edge red shifts to 800 nm. That is, crystalline $Sb_2S_3$ is transparent for wavelengths longer than 800 nm.

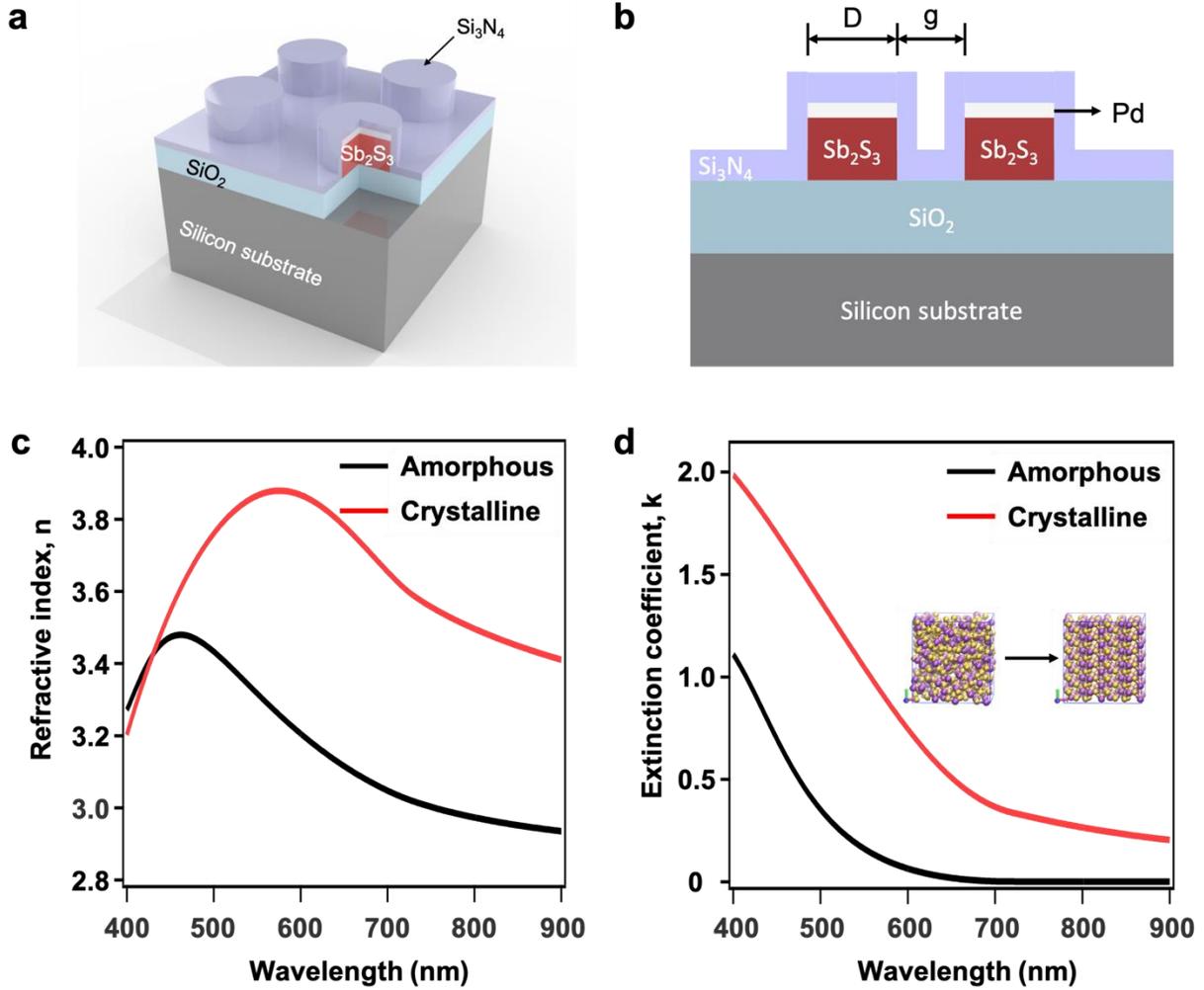

**Figure 1. Schematic of dielectric $Sb_2S_3$ antenna array and refractive index information.** (a) Schematic design of the $Sb_2S_3$ nanodisk array on Si substrate with 200-nm-thick $SiO_2$ film and being protected by 70-nm-thick $Si_3N_4$ film. (b) Schematic of nanostructure cross-section. The diameter of the nanoantenna and the gap between the nanoantennas are denoted as $D$ and $g$ respectively. (c) Refractive index $n$ and (d) Extinction coefficient $k$ of $Sb_2S_3$ at both amorphous and crystalline states.

High resolution color prints that exploit Mie scattering from dielectric nanoparticles can be used in both transmission and reflection modes using either forward or backward scattering; that is by either exploiting Kerker's 1$^{st}$ or 2$^{nd}$ conditions. Here, the nanostructured $Sb_2S_3$ color palette was designed for reflective color using the 2$^{nd}$ Kerker's condition. We designed the color resonators by performing 3D full-wave simulations of resonator arrays using the finite-difference time-domain (FDTD) method. $Sb_2S_3$ nanodiscs with a diameter $D$ of 190 nm and a gap size $g$ of 90 nm show a large color change between the amorphous and crystalline states, as shown in Figure 2. For instance, at the amorphous state, the simulated reflection spectra show a resonance dip at 565 nm that red shifts to 630 nm after crystallization. At the dip wavelength of 565 nm, the reflectance is changed by ~10-fold, when the material is switched from amorphous to crystalline state (see the detailed comparison at Fig. S4). The simulated spatial distributions of electric field |$E$| and magnetic field |$H$| at the resonance wavelengths for both amorphous and crystalline states are shown in Figure 2(c)-(d) and Figure 2(g)-(h), respectively. In the amorphous state, the simulated reflection spectrum shows a sharp resonance, and the electric field is highly confined in the dielectric nanoantenna. After annealing, the material switches into the more absorptive crystalline state that broadens the resonance. The electric field distribution also shows weak optical confinement to the nanodisc, especially at the shorter wavelengths, such as 425 nm as shown in Figure 2(h).

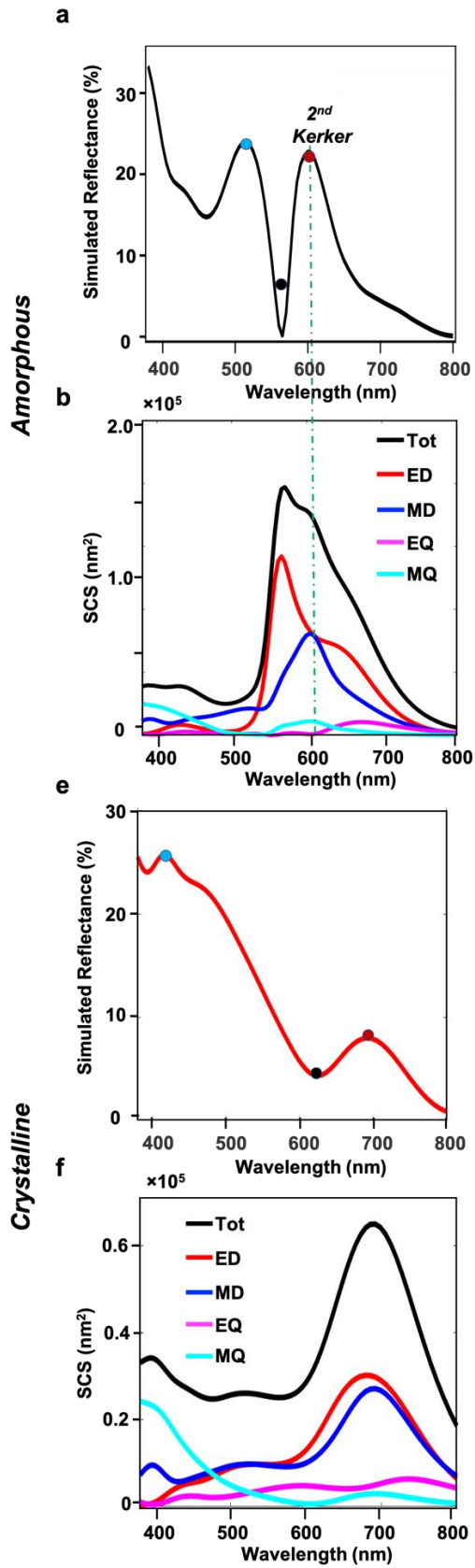
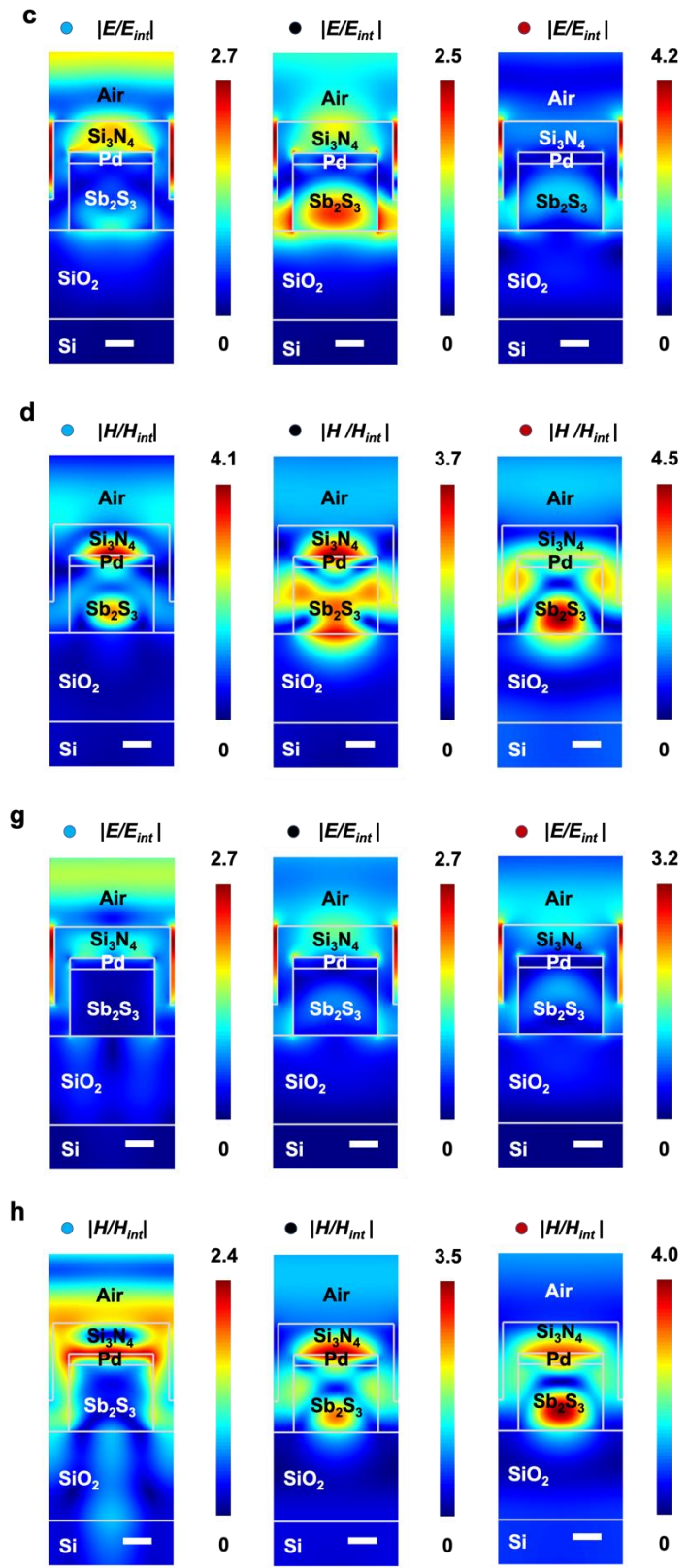

**Figure 2. Finite-difference time-domain (FDTD) simulation of the nanostructured Sb₂S₃ nanoantenna array with a diameter *D* of 190 nm and a gap size *g* of 90 nm.** Reflectance spectra and multipolar decomposition calculations of the scattering cross sections (SCS), $\sigma_{scattering}$, for the *MD*, *ED*, *MQ*, and *EQ* modes for the amorphous state (a)-(b) and the crystalline state (e)-(f). (c)-(d) Simulated electric and magnetic field distributions for amorphous state at the wavelengths of 515 nm, 565 nm, and 600 nm. (g)-(h) Simulated electric and magnetic field distributions for crystalline state at the wavelengths of 425 nm, 630 nm and 690 nm. The scale bars in all the mode pattern plots denote 100 nm.

In order to shed light on the optical response of the structure, multipole decomposition was carried out on Sb₂S₃ nanopillars according to the detailed formula as reported in the reference.[3] The scattering cross-section was decomposed into electric dipole (*ED*), magnetic dipole (*MD*), electric quadrupole (*EQ*), and magnetic quadrupole (*MQ*) modes as plotted in Figure 2(b) and Figure 2(f). By analyzing the multipole modes in Figure 2(b), it can be observed that the absorption dip around 565 nm in Figure 2(a) is due to excitation of a strong *ED* while the reflectance peak around 600 nm is due to *MD* excitation. The strong *ED* excitation in pillars can be observed from the |*E*| field distribution in Figure 2(c) as well. The reflectance peak around 600 nm is approximately due to fulfillment of the 2$^{nd}$ Kerker's condition, *i.e.*, phase(*MD*) - phase(*ED*) = π and abs(*MD*) = abs(*ED*) as highlighted in Figure S5. Thus, when the Sb₂S₃ nanodisc is switched from amorphous to crystalline, a large color change is expected due to the changes of the excited electric and magnetic dipole resonances. The 2$^{nd}$ Kerker's condition for Sb₂S₃ nanoantenna with the same geometry parameters are no longer satisfied when Sb₂S₃ is changed to crystalline state.

Figure 3(a) shows the bright-field optical microscope images for a matrix of 15×15 μm² pixels in the amorphous state. Note, each color pixel is composed of an array of Sb₂S₃ nanodisc

Mie resonators. The diameter $D$ of the resonators increases from 50 nm to 280 nm, while the gap size $g$ is increased from 10 nm to 200 nm. As one would expect, the color of the array strongly depends on the disc diameter and the gap. After the nano-resonators were crystallized at 300 °C, a radical color change was observed as shown in Figure 3(b). For example, nanoantennas with diameter than *D>140* nm became blue and darker. It is because that $Sb_2S_3$ nanodiscs at the amorphous state can support resonances in the wavelength between 600 nm and 800 nm, but crystallization causes the optical band gap energy to be lower, causing higher absorption for wavelengths longer than 600 nm. This increase in absorption is clearly shown in the extinction coefficient, $k$ in Figure 1(d).

In order to show the corresponding spectrum shift, Figure 3(c) and Figure 3(d) present the reflection spectra for $Sb_2S_3$ nanoantenna arrays with a gap size $g$ of 90 nm and the diameter $D$ being changed from 70 nm to 280 nm, for both amorphous and crystalline states. These spectrum shifts are shown *via* the corresponding obvious color change as well. Moreover, for both amorphous and crystalline states, the reflection spectra resonances exhibit a red shift when the diameter increases. This is because that the electric and magnetic dipole resonances depend on the size of the nanoantenna.

The SEM image of the nanodisc array ($D=190$ nm & $g=90$ nm) before and after depositing $Si_3N_4$ are shown in Figure 3(e) and (f) respectively. Note that the gap size g in Figure 3(e) is less than 90 nm and its diameter of the nanodisc is larger than 190 nm, due to the deposition of $Si_3N_4$ after the etching process. The bright field optical microscope image of $Sb_2S_3$ color palette and the SEM image of nanoantenna array before $Si_3N_4$ deposition is shown in the supporting information Figure S6. Crystallization causes a substantial resonance shift shown in Figure 3(h), which compares the measured reflection spectra of the as-deposited amorphous and crystallized states.

We see that the resonance dip wavelength shifts from 532 nm in the amorphous state to 659 nm in the crystalline state. The reflected color changes from pink to dark blue, as shown by the International Commission on Illumination (CIE) 1931 chromaticity diagram in Figure 3(g). To further confirm the phase transition of $Sb_2S_3$ due to the annealing process, the Raman spectra were measured and shown in Figure 3(i). In the amorphous state, there are two broadband at 100 $cm^{-1}$ and 290 $cm^{-1}$. After annealing, there are several peaks at 156 $cm^{-1}$, 184 $cm^{-1}$, 238 $cm^{-1}$, 284 $cm^{-1}$ and 309 $cm^{-1}$ corresponding to the $Sb_2S_3$ in the orthorhombic phase. Thus, this Raman measurement indeed confirms the $Sb_2S_3$ phase transition from amorphous to crystalline state. The simulated $Sb_2S_3$ color palette and reflection spectra of $Sb_2S_3$ nanoantenna arrays for both amorphous and crystalline states can be found in Figure S7.

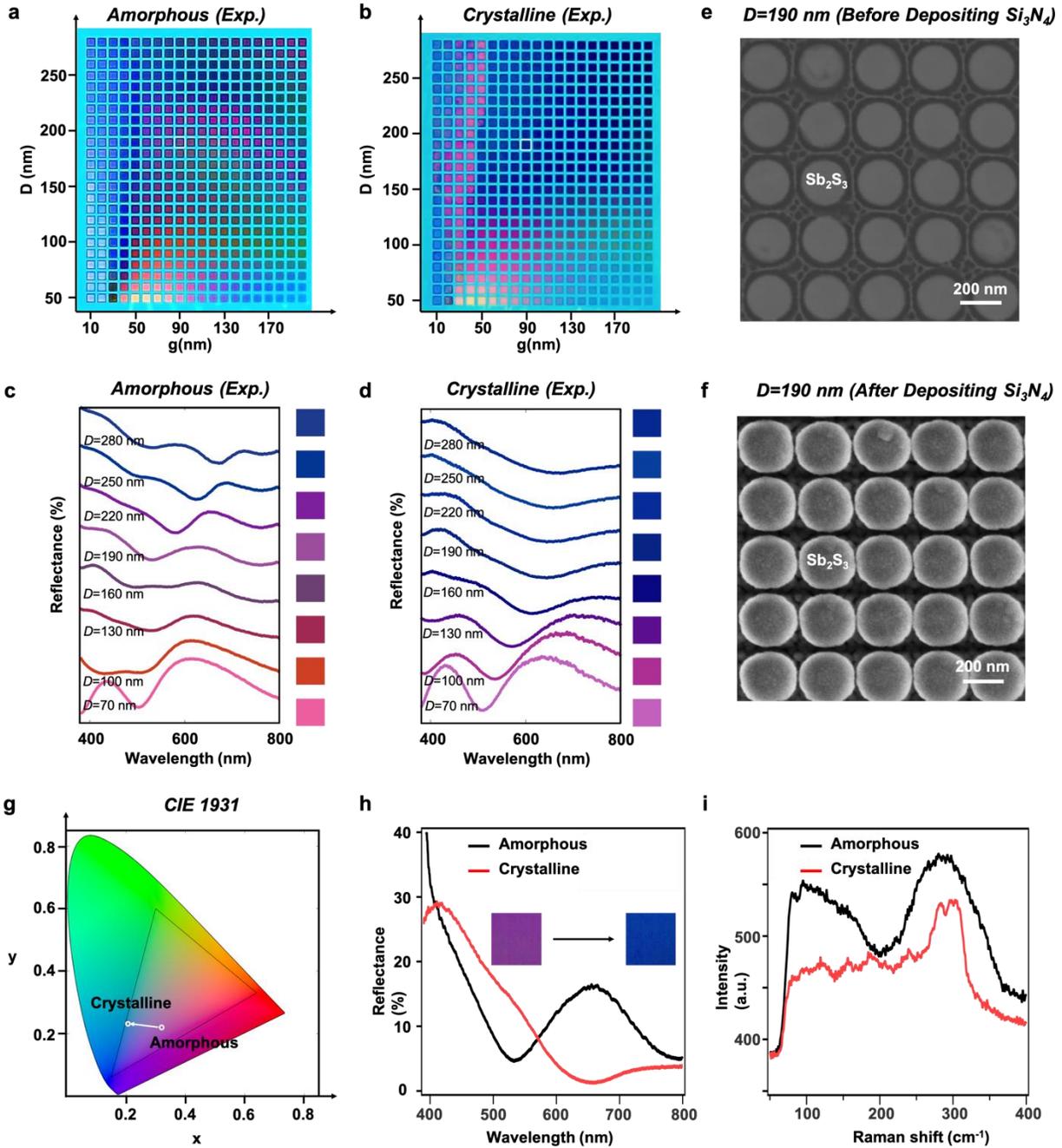

**Figure 3. Reflected color of the nanostructured Sb$_2$S$_3$ nanoantenna at amorphous and crystalline states.** Bright field optical microscope images of the nanostructured color palette when Sb$_2$S$_3$ nanostructures are in the (a) amorphous state and (b) after annealing at 300 °C for one hour in the Argon atmosphere. Here, the dash square line highlights the Sb$_2$S$_3$ nanoantenna array with a diameter $D$ of 190 nm and a gap size $g$ of 90 nm. (c)-(d) Reflection spectra for nanoantenna with gap size *g* of 90 nm, diameter *D* being changed from 70 nm to 280 nm for amorphous and crystalline states respectively. (e)-(f) Scanning electron micrograph (SEM) images of $Sb_2S_3$ nanoantenna array with a gap of 90 nm and diameter of 190 nm before and after depositing $Si_3N_4$. (g) Corresponding International Commission on Illumination (CIE) 1931 chromaticity diagram for nanoantenna in the amorphous state and after annealing. (h) Direct comparison of the reflection spectra for nanoantenna in the amorphous state and crystalline state. (i) Measured Raman spectra at the amorphous and crystalline states.

For both measured and simulated reflection spectra, the resonances exhibit a red shift when $Sb_2S_3$ changes from amorphous state to crystalline state. This is due to the refractive index increment when there is structural phase transition in $Sb_2S_3$. In addition, the resonance in the crystalline state is weaker than that in the amorphous state. This is due to the Mie resonances cannot be excited efficiently when there is much higher loss at the crystalline state. The scattering efficiency of *ED* and *MD* for $Sb_2S_3$ nanoantenna arrays are shown in Figure S8 for both states. The single $Sb_2S_3$ nanoantenna element exhibited a similar performance as shown in Figure S9. It shows that the color change results from the Mie resonance change when $Sb_2S_3$ changes from amorphous to crystalline state.

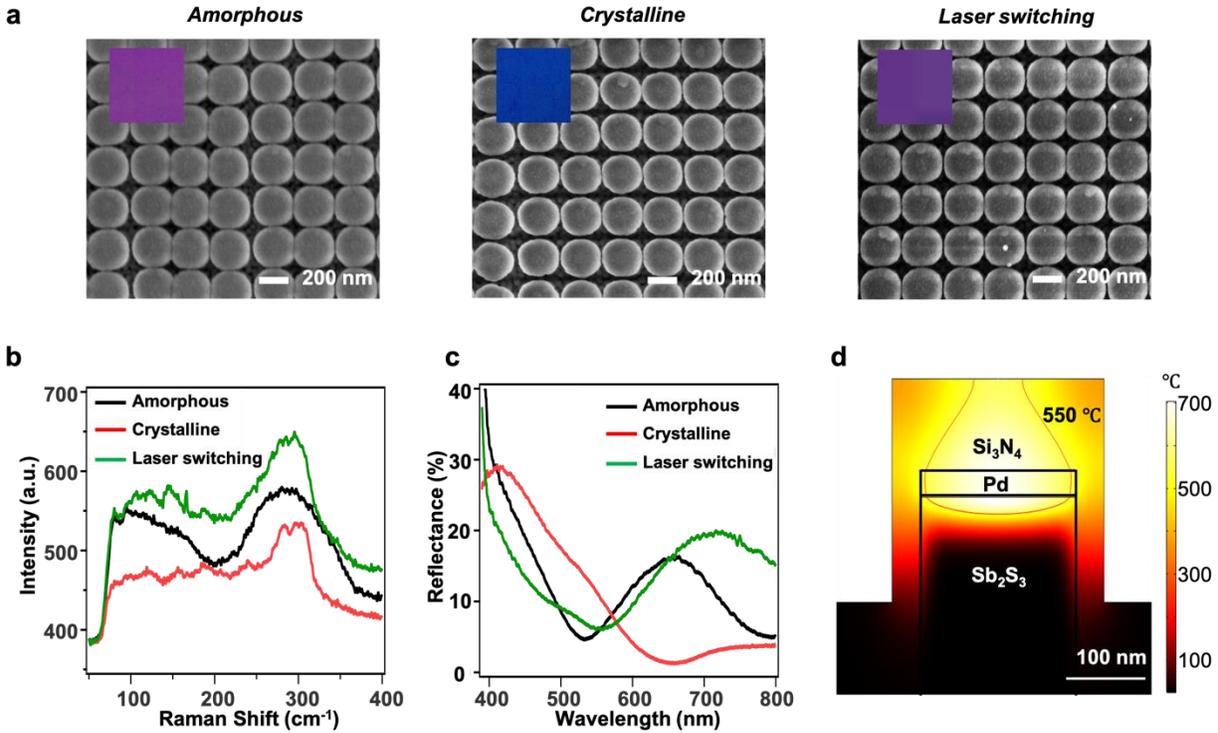

**Figure 4. Characterization results of the tunable Sb$_2$S$_3$ dielectric nanodisc array.** (a) SEM images for Sb$_2$S$_3$ nanodiscs with a diameter *D* of 190 nm and a gap size g of 90 nm in the amorphous state, after annealing and after laser amorphization. The corresponding optical micrographs are shown in the insets. (b) Raman spectra and (c) Reflection spectra for Sb$_2$S$_3$ nanodisc arrays in the amorphous state, after annealing and after laser amorphization. (d) Thermal simulation for Sb$_2$S$_3$ nanoantenna laser amorphization using femtosecond laser.

Thus far, we have shown using simulations and experiments that the nanodisc arrays can exhibit substantial changes to their color when they are annealed above the crystallization. At the same time, there exists some slight discrepancies between the experiment results and simulated results, due to the following two reasons. One reason is that the thickness of 20-nm-thick nanostructured Pd hard mask will be changed after etching, and it is hard to measure the exact thickness value, and this thickness of Pd hard mask may affect the simulation accuracy. Another reason is that the shape of nanoantenna is no longer a perfect cylinder after Si$_3$N$_4$ deposition *via*

ICP CVD process as shown in Figure 3(f), and it could affect the resonance wavelengths of the reflection spectra. Now, we show that reversible changes to the scattered color can be induced using laser pulses. Again, we see that the reflected color from the nanodisc arrays changes from purple to blue when the $Sb_2S_3$ nanodisc array changes state. To tune the resonant wavelength of the $Sb_2S_3$ dielectric nanodiscs, we used a train of 780 nm, 100-fs laser pulses at a frequency of 80 MHz. These pulses were raster scanned across the nanodisc arrays. The optimal scan speed was 10 μm/s, and the average laser power was 12 mW. Figure 4(a) shows the SEM images and optical micrographs for $Sb_2S_3$ nanodisc array ($D$=190 nm & $g$=90 nm) in the amorphous state, after annealing, and after laser amorphization. The reflected color of nanodisc arrays change from purple to blue when $Sb_2S_3$ changes from the amorphous to the crystalline state. After laser amorphization, the nanoantenna sample switches back to the amorphous purple color. The SEM images for different states show that the nanodiscs are not damaged by this crystallization and laser amorphization cycle. The 780 nm wavelength was chosen because the absorption of $Sb_2S_3$ in the crystalline state is sufficiently low that the nanodisc can be heated above the melting temperature of $Sb_2S_3$ (550 °C) as shown in Figure 4(d). In this way, $Sb_2S_3$ can be reamorphized *via* melt-quench process. This result is important because reversibility of color is necessary for displays and active filters.

The amorphous state of the laser switched state was confirmed using Raman spectroscopy. The amorphous $Sb_2S_3$ spectrum shows broad peaks at 100 $cm^{-1}$ and 300 $cm^{-1}$. We see in Figure 4(b) that the Raman peaks are substantially broadened at 100 $cm^{-1}$ and 300 $cm^{-1}$, where the crystalline features at 156 $cm^{-1}$ and 184 $cm^{-1}$ are heavily suppressed. We also see that these features are corresponding to changes in the reflection spectra resonance being blue-shifted from 566 nm in the crystalline state to 532 nm in the amorphous state as shown in Figure 4(c).

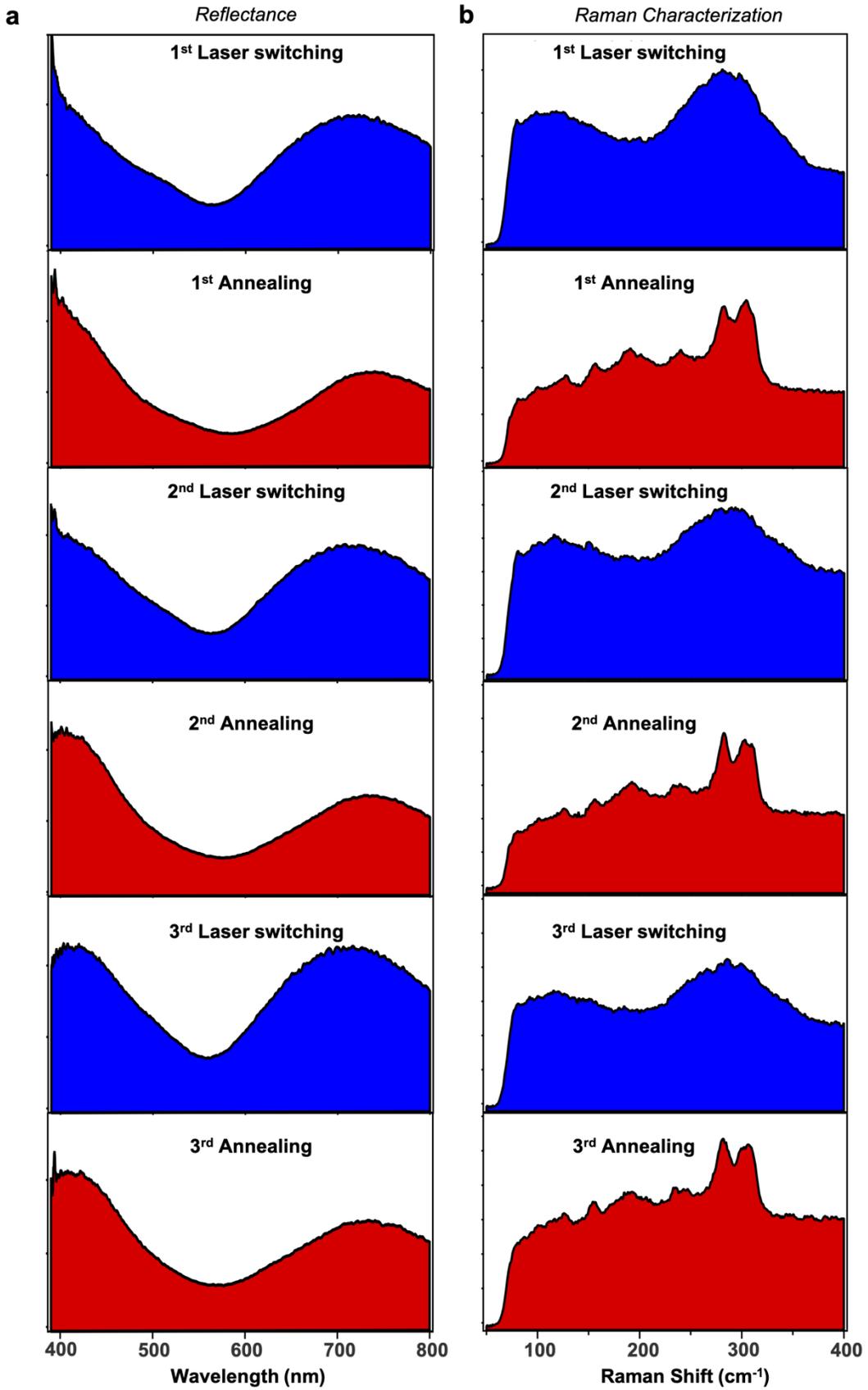

**Figure 5. Cyclability test of the Sb$_2$S$_3$ nanodisc arrays.** (a) Reflectance spectra for laser amorphization and after annealing states. (b) Raman spectra for laser amorphization and after annealing states.

The cyclability of the color change is important for display applications. To study the ability of cyclability of our designed Sb$_2$S$_3$ nanodisc array, we used the same parameters for laser amorphization and hotplate annealing test at 300 °C several times. The resonance positions of reflection spectra for both amorphous and crystalline states do not show deterioration during the cyclability test, as shown in Figure 5(a). We also measured the Raman spectrum after each process to ensure that the material structural state is correct and the material is not damaged, as shown in Figure 5(b). After each laser amorphization process, the Sb$_2$S$_3$ nanoantenna arrays exhibit the amorphous state signal, and it will be in the crystalline state after the annealing process.

The cyclability test shows that the reflected spectrum can be switched between the two states. However, the reflection spectrum of the crystallized nanodisc array is slightly different before and after 1$^{st}$ laser amorphization. It is due to the partial amorphization, because the thickness of Sb$_2$S$_3$ is 150 nm, and only the top part of the material can reach melting temperature to achieve amorphization, see Figure 4(d). In this way, we can achieve multilevel switching to show continuous analogue changes in colors on the same nanoantenna arrays, *via* carefully engineering the laser power and scanning speed. If the laser amorphization process can be conducted using laser at other wavelength and at a lower temperature to increase the quenching rate of Sb$_2$S$_3$ in the nanostructure, we expect that it is possible to achieve the same reflection spectrum as the as-deposited state, which will be further explored in the future.

**Conclusion**

In this paper, we have demonstrated a way to tune Mie resonances in the visible spectrum. We believe that this is the first such demonstration with reversibility. Arrays of $Sb_2S_3$ nanodiscs with Mie resonances exhibit with strong structural color, which can be tuned via heat induced crystallization. The color was reversed by amorphization with laser pulses. This demonstrates that the Mie resonance can be manipulated in the visible spectrum via amorphous to crystalline phase transitions. Under the condition of resonant absorption, changes in the optical band gap produced the greatest changes in color. These results show that Mie resonances at visible regime are not only possible, but can potentially enable a wide variety of tunable nanophotonic devices, such as high-resolution color displays,[37] holographic displays[38] and miniature LiDAR scanning systems.[18,19]

## METHODS

**$Sb_2S_3$ Sputtering via Radio Frequency (RF) Sputtering.** $Sb_2S_3$ film with a thickness of 150 nm was first deposited onto the silicon substrate with 200-nm-thick $SiO_2$, via RF sputtering. The chamber base pressure was $5.7 \times 10^{-5}$ Pa, and the sputtering pressure was 0.5 Pa. The 150 nm thin film was deposited from a two inches $Sb_2S_3$ alloy target with a purity of 99.9%.

**Nanopatterning of PMMA Resist by Electron Beam Lithography (EBL).** First, the poly-methyl methacrylate (PMMA, A5) resist was diluted in anisole solvent with a ratio of 1:1, where the resist was spin coated onto a cleaned substrate at 3k round-per-minute (rpm) to obtain an PMMA thickness of ~100 nm. After that, the PMMA resist is baked at 180 °C for 2 minutes to evaporate the anisole solvent. Electron beam exposure was then carried out with an acceleration voltage of 100 keV and a beam current of 100 pA, with an exposure dosage of ~5 $mC/cm^2$. After the electron beam exposure, the PMMA sample was developed in the methyl isobutyl ketone (MIBK) and isopropyl alcohol (IPA) solution with a volume mixing ratio of 1:3, at a temperature of -10 °C for 15 seconds,[39] followed by the blow dry of the sample via nitrogen gas flow.

**Evaporation of Pd Film, Lift Off Process, and Dry Etching of $Sb_2S_3$ and Growth of $Si_3N_4$ Film by ICP-CVD.** 20-nm-thick Pd film with 2-nm Ti adhesion layer was evaporated onto the nanostructured PMMA resist by using electron beam evaporator (Denton), at a pressure of $5 \times 10^{-7}$ torr and a deposition rate of 0.5 Å/s. The lift off process of this sample was carried out in the N-Methyl-2-pyrrolidone (NMP) solution at the temperature of 70 °C. After the lift off process, the sample is cleaned by Aceton solvent, IPA solvent and followed by nitrogen gun. Then, $Sb_2S_3$ etching was carried out by inductively-coupled-plasma (ICP, Oxford Instruments Plasmalab System 100) with the $Cl_2$ gas chemistry,[40] at a DC power of 100 watts, a coil power of 300 watts, $Cl_2$ with a flow rate of 22 sccm (standard-cubic-centimeters-per-minute), under a process pressure of 10 mTorr, and a temperature of 10 °C. To protect the $Sb_2S_3$ nanodiscs during the annealing process, 70-nm-thick $Si_3N_4$ film was grown on top of the nanodisc array by using ICP chemical vapor decomposition (CVD) at 150 °C.

**Annealing and Laser Amorphization Process.** Since these sample exhibit a strong colour change in the visible spectrum, the crystallisation temperature is readily observed as a colour change. This might not be so simple for other nano disc array devices that do not exhibit optical changes in the visible spectrum. Annealing the sample in a Linkam furnace at 300 °C for 1 hour was sufficient to fully crystallise the nanodiscs. The annealing was performed in a flowing argon atmosphere (4 SCCM) to prevent oxidation, with a heating rate of 5 °C/min. The laser amorphization for $Sb_2S_3$ nanodiscs was conducted by using femtosecond laser pluses from Photonics Professional GT nanoscale 3D printing system by Nanoscribe GmbH. The femtosecond laser pulses are at the wavelength 780 nm with 80-MHz repetition rate and 100-fs pulse width. The raster scan was performed on the nanodisc array sample. The optimized scan speed was 10 μm/s, and the laser power was 12 mW.

**Characterizations.** The optical reflectance spectra of $Sb_2S_3$ nanodisc arrays were measured by using Craic micro-spectrometer, with a ×36 objective lens with a numerical aperture of 0.50. Moreover, the optical images of the color palettes were captured by the Olympus microscope (MX61) with a software of "analySIS", where the objective lens used is ×10 (MPlanFL N, NA=0.30). Before the image capture, a white color balancing was first carried out on a 100-nm-thick aluminum film, which was prepared by electron beam evaporator.[41] SEM images were taken at an acceleration voltage of 10 keV with the SEM model number of Elionix, ESM-9000.

**Numerical Simulations.** The optical simulations of the $Sb_2S_3$ nanodisc arrays were performed using a commercial finite-difference time domain software, FDTD Solutions (Lumerical Solutions, Inc.). A plane source with wavelengths between 380 nm and 800 nm was input in the negative z-direction perpendicular to the nanodisc array and substrate. A field monitor was placed 100 nm above the plane source to measure the reflected power, while another field monitor was placed perpendicular to the substrate along $x$=0 nm to measure the electric and magnetic fields in the cross section of the nanodisc. Periodic boundary conditions were set for the $x$- and $y$-boundaries, while the perfectly matched layer (PML) boundary condition was set for the $z$-boundaries. A mesh override region with a mesh size of 2 nm was drawn surrounding the $Sb_2S_3$/Pd nanodisc, $Si_3N_4$ coating and $SiO_2$ layer to increase accuracy. Refractive index data for Si, $SiO_2$ and Pd were taken from Palik,[42] while refractive index data for $Si_3N_4$ were taken from Philipp.[43] Refractive index data for the amorphous and crystalline phases of $Sb_2S_3$ were measured using an ellipsometer as shown in Figure 1. Multipole decomposition analysis was carried out using a Finite Element Method based software (COMSOL Multiphysics 5.6, Optical Wave Optics Module). A single unit-cell of the structure was simulated using periodic boundary conditions under normally incident plane-wave. Calculated fields were used to perform the multipole decomposition based on the formula as shown in the supporting information of Reference.[3]

**Thermal simulation.** The heat generation by femtosecond laser pulses can be regarded as an instantaneous process due to the extreme short duration time of pulses. Then, the heat gradually diffuses outward. The energy flux of a femtosecond laser pulse is written as

$$J(r) = \frac{2q}{\pi w_0^2} \exp\left(-2\frac{r^2}{w_0^2}\right),$$

where $q$ is the pulse energy, $w_0$ is the $1/e^2$ radius, $r$ is the spatial coordinate. Governed by the Beer-Lambert absorption law, the intensity of the femtosecond pulse decreases with the propagation in material. The absorbed pulse energy density is

$$\Delta U(r,z) = (1-R) \cdot J(r) \cdot \alpha \cdot \exp(-\alpha z),$$

where $R$ is the reflectivity of sample, $\alpha$ is the absorption coefficient of material. Using the heat capacity equation, the temperature variation can be calculated by

$$\Delta T(r,z) = \frac{\Delta U(r,z)}{\rho c} = \frac{(1-R)\alpha}{\rho c} \frac{2q}{\pi w_0^2} \exp\left(-2\frac{r^2}{w_0^2}\right) \exp(-\alpha z),$$

where $\rho$ is the mass density and $c$ is the specific heat capacity.

Now, we have obtained the instantaneous temperature distribution $\Delta T(r,z)$ generated by a femtosecond laser pulse. $\Delta T(r,z)$ will be used to the heat diffusion process as initial temperature condition, and it governed by unsteady heat conduction equation:

$$\rho c \frac{\partial T(r,z,t)}{\partial t} = \nabla \cdot \left( k \nabla \cdot T(r,z,t) \right),$$

where $k$ is the heat conduction. Due to a layer of Pd on the $Sb_2S_3$, the absorption of femtosecond pulses is brought by Pd.

## ASSOCIATED CONTENTS

**Supporting Information**

Reflection spectra for flat substrate, Nanofabrication process of $Sb_2S_3$ nanoantenna array, $Sb_2S_3$ nanoantenna color palette for amorphous and crystalline state in CIE chromaticity diagram, Reflection spectra, excited multipoles magnitude and corresponding phase in for $Sb_2S_3$ nanoantenna with D = 190 nm, gap = 90 nm in the amorphous and crystalline phase, $Sb_2S_3$ color palette and SEM image for $Sb_2S_3$ nanoantenna array with D = 190 nm and gap = 90 nm before depositing $Si_3N_4$, Simulated $Sb_2S_3$ color palette and reflection spectra for nanoantenna with gap = 90 nm, diameter from 70 nm to 280 nm for amorphous and crystalline states, scattering cross-section of $Sb_2S_3$ nanoantenna arrays with, gap = 90 nm, diameter varied from 70 nm to 280 nm, scattering cross-section of $Sb_2S_3$ nanoantenna single element with diameter varied from 70 nm to 280 nm.

## AUTHOR INFORMATION

**Corresponding Author**

*E-mail: robert_simpson@sutd.edu.sg.

*E-mail: joel_yang@sutd.edu.sg.

*E-mail: dongz@imre.a-star.edu.sg.

**ORCID**

Zhaogang Dong: 0000-0002-0929-7723

Joel K. W. Yang: 0000-0003-3301-1040

**Author contributions**

L.L., Z.D., J.K.W.Y. and R.E.S. conceived the concept, designed the experiments and wrote the manuscript. L.L. did the sputtering of $Sb_2S_3$ films, annealing of $Sb_2S_3$ nanostructures, and performed the laser amorphisation and Raman spectroscopy measurements. Z.D. did the electron beam lithography and developed the whole fabrication processes. Z.D. and F.T. performed the deposition of Pd film by electron-beam evaporator, lift off process, dry etching and ICP CVD growth of $Si_3N_4$ film layer to protect the nanostructured $Sb_2S_3$. R.J.H.N. performed the FDTD simulations. H.W. performed the initial laser induced amorphisation of $Sb_2S_3$ nanostructures. S.D.R. performed the multipole decomposition analysis and FDTD simulations. Y.W. did the thermal simulation and analysis. H.S.L. and L.L. did the optical reflectance measurements. All authors analyzed the data, read and corrected the manuscript before the submission.

**Notes**

The authors declare no competing financial interests.

**ACKNOWLEDGMENTS**

This work is supported by the funding support of A*STAR AME IRG project with a project number of A20E5c0093. In addition, R.E.S. would like to acknowledge the funding supporting from the Nano Spatial Light Modulators (NSLM) AME grant (A18A7b0058). Z.D. also would like to acknowledges the funding support by A*STAR career development award (CDA) with the grant number of 202D8088.

**REFERENCES**

1. Kuznetsov, A. I.; Miroshnichenko, A. E.; Fu, Y. H.; Zhang, J.; Luk'yanchuk, B., Magnetic light. *Sci. Rep.* **2012,** *2*, 492.
2. Evlyukhin, A. B.; Novikov, S. M.; Zywietz, U.; Eriksen, R. L.; Reinhardt, C.; Bozhevolnyi, S. I.; Chichkov, B. N., Demonstration of Magnetic Dipole Resonances of Dielectric Nanospheres in the Visible Region. *Nano Lett.* **2012,** *12* (7), 3749-3755.
3. Dong, Z.; Ho, J.; Yu, Y. F.; Fu, Y. H.; Paniagua-Dominguez, R.; Wang, S.; Kuznetsov, A. I.; Yang, J. K. W., Printing Beyond sRGB Color Gamut by Mimicking Silicon Nanostructures in Free-Space. *Nano Lett.* **2017,** *17* (12), 7620-7628.
4. Kerker, M.; Wang, D. S.; Giles, C. L., Electromagnetic scattering by magnetic spheres. *J. Opt. Soc. Am.* **1983,** *73* (6), 765-767.
5. Fu, Y. H.; Kuznetsov, A. I.; Miroshnichenko, A. E.; Yu, Y. F.; Luk'yanchuk, B., Directional visible light scattering by silicon nanoparticles. *Nat. Commun.* **2013,** *4*, 1527.
6. Jin, L.; Dong, Z.; Mei, S.; Yu, Y. F.; Wei, Z.; Pan, Z.; Rezaei, S. D.; Li, X.; Kuznetsov, A. I.; Kivshar, Y. S.; Yang, J. K. W.; Qiu, C.-W., Noninterleaved Metasurface for $(2^{6}-1)$ Spin- and Wavelength-Encoded Holograms. *Nano Lett.* **2018,** *18* (12), 8016-8024.
7. Qu, G.; Yang, W.; Song, Q.; Liu, Y.; Qiu, C.-W.; Han, J.; Tsai, D.-P.; Xiao, S., Reprogrammable meta-hologram for optical encryption. *Nat. Commun.* **2020,** *11* (1), 5484.
8. Huang, K.; Dong, Z.; Mei, S.; Zhang, L.; Liu, Y.; Liu, H.; Zhu, H.; Teng, J.; Luk'yanchuk, B.; Yang, J. K. W.; Qiu, C.-W., Silicon multi-meta-holograms for the broadband visible light. *Laser Photon. Rev.* **2016,** *10* (3), 500-509.
9. Yang, W.; Xiao, S.; Song, Q.; Liu, Y.; Wu, Y.; Wang, S.; Yu, J.; Han, J.; Tsai, D.-P., All-dielectric metasurface for high-performance structural color. *Nat. Commun.* **2020,** *11* (1), 1864.
10. Zhu, X.; Yan, W.; Levy, U.; Mortensen, N. A.; Kristensen, A., Resonant laser printing of structural colors on high-index dielectric metasurfaces. *Sci. Adv.* **2017,** *3* (5), e1602487.
11. Dong, Z.; Jin, L.; Rezaei, S. D.; Wang, H.; Chen, Y.; Tjiptoharsono, F.; Ho, J.; Gorelik, S.; Ng, R. J. H.; Ruan, Q.; Qiu, C.-W.; Yang, J. K. W., Schrödinger's Red Pixel by Quasi Bound-State-In-Continuum. *arXiv* **2021,** *2106.12285*.
12. Ha, S. T.; Fu, Y. H.; Emani, N. K.; Pan, Z.; Bakker, R. M.; Paniagua-Domínguez, R.; Kuznetsov, A. I., Directional lasing in resonant semiconductor nanoantenna arrays. *Nat. Nano.* **2018,** *13* (11), 1042-1047.

13. Huang, C.; Zhang, C.; Xiao, S.; Wang, Y.; Fan, Y.; Liu, Y.; Zhang, N.; Qu, G.; Ji, H.; Han, J.; Ge, L.; Kivshar, Y.; Song, Q., Ultrafast control of vortex microlasers. *Science* **2020,** *367* (6481), 1018-1021.
14. Dong, Z.; Gorelik, S.; Paniagua-Dominguez, R.; Yik, J.; Ho, J.; Tjiptoharsono, F.; Lassalle, E.; Rezaei, S. D.; Neo, D. C. J.; Bai, P.; Kuznetsov, A. I.; Yang, J. K. W., Silicon Nanoantenna Mix Arrays for a Trifecta of Quantum Emitter Enhancements. *Nano Lett.* **2021,** *21* (11), 4853–4860.
15. Ho, J.; Fu, Y. H.; Dong, Z.; Paniagua-Dominguez, R.; Koay, E. H. H.; Yu, Y. F.; Valuckas, V.; Kuznetsov, A. I.; Yang, J. K. W., Highly Directive Hybrid Metal–Dielectric Yagi-Uda Nanoantennas. *ACS Nano* **2018,** *12* (8), 8616-8624.
16. Koshelev, K.; Lepeshov, S.; Liu, M.; Bogdanov, A.; Kivshar, Y., Asymmetric Metasurfaces with High-Q Resonances Governed by Bound States in the Continuum. *Physical Review Letters* **2018,** *121* (19), 193903.
17. Dong, Z.; Mahfoud, Z.; Paniagua-Domínguez, R.; Wang, H.; Fernández-Domínguez, A. I.; Gorelik, S.; Ha, S. T.; Tjiptoharsono, F.; Kuznetsov, A. I.; Bosman, M.; Yang, J. K. W., Photonic Bound-States-in-the-Continuum Observed with an Electron Nanoprobe. *ArXiv* **2021**, 2105.04220.
18. Li, S.-Q.; Xu, X.; Maruthiyodan Veetil, R.; Valuckas, V.; Paniagua-Domínguez, R.; Kuznetsov, A. I., Phase-only transmissive spatial light modulator based on tunable dielectric metasurface. *Science* **2019,** *364* (6445), 1087-1090.
19. Park, J.; Jeong, B. G.; Kim, S. I.; Lee, D.; Kim, J.; Shin, C.; Lee, C. B.; Otsuka, T.; Kyoung, J.; Kim, S.; Yang, K.-Y.; Park, Y.-Y.; Lee, J.; Hwang, I.; Jang, J.; Song, S. H.; Brongersma, M. L.; Ha, K.; Hwang, S.-W.; Choo, H.; Choi, B. L., All-solid-state spatial light modulator with independent phase and amplitude control for three-dimensional LiDAR applications. *Nat. Nano.* **2021,** *16* (1), 69-76.
20. Kuznetsov, A. I.; Miroshnichenko, A. E.; Brongersma, M. L.; Kivshar, Y. S.; Luk'yanchuk, B., Optically resonant dielectric nanostructures. *Science* **2016,** *354* (6314), aag2472.
21. Sautter, J.; Staude, I.; Decker, M.; Rusak, E.; Neshev, D. N.; Brener, I.; Kivshar, Y. S., Active Tuning of All-Dielectric Metasurfaces. *ACS Nano* **2015,** *9* (4), 4308-4315.
22. Komar, A.; Fang, Z.; Bohn, J.; Sautter, J.; Decker, M.; Miroshnichenko, A.; Pertsch, T.; Brener, I.; Kivshar, Y. S.; Staude, I.; Neshev, D. N., Electrically tunable all-dielectric optical metasurfaces based on liquid crystals. *Applied Physics Letters* **2017,** *110* (7), 071109.
23. Hashemi, M. R. M.; Yang, S.-H.; Wang, T.; Sepúlveda, N.; Jarrahi, M., Electronically-Controlled Beam-Steering through Vanadium Dioxide Metasurfaces. *Sci. Rep.* **2016,** *6* (1), 35439.
24. Wang, Q.; Rogers, E. T. F.; Gholipour, B.; Wang, C.-M.; Yuan, G.; Teng, J.; Zheludev, N. I., Optically reconfigurable metasurfaces and photonic devices based on phase change materials. *Nat. Photon.* **2016,** *10* (1), 60-65.
25. Loke, D.; Lee, T. H.; Wang, W. J.; Shi, L. P.; Zhao, R.; Yeo, Y. C.; Chong, T. C.; Elliott, S. R., Breaking the Speed Limits of Phase-Change Memory. *Science* **2012,** *336* (6088), 1566-1569.
26. Wuttig, M.; Bhaskaran, H.; Taubner, T., Phase-change materials for non-volatile photonic applications. *Nat. Photon.* **2017,** *11* (8), 465-476.
27. Shportko, K.; Kremers, S.; Woda, M.; Lencer, D.; Robertson, J.; Wuttig, M., Resonant bonding in crystalline phase-change materials. *Nat. Mater.* **2008,** *7* (8), 653-658.

28. Wang, Y.; Landreman, P.; Schoen, D.; Okabe, K.; Marshall, A.; Celano, U.; Wong, H. S. P.; Park, J.; Brongersma, M. L., Electrical tuning of phase-change antennas and metasurfaces. *Nat. Nano.* **2021**.
29. Abdelraouf, O. A.; Anthur, A. P.; Dong, Z.; Liu, H.; Wang, Q.; Krivitsky, L.; Wang, X. R.; Wang, Q. J.; Liu, H., Multistate Tuning of Third Harmonic Generation in Fano-Resonant Hybrid Dielectric Metasurfaces. *arXiv* **2021**, *2105.04233*.
30. Zhang, Y.; Fowler, C.; Liang, J.; Azhar, B.; Shalaginov, M. Y.; Deckoff-Jones, S.; An, S.; Chou, J. B.; Roberts, C. M.; Liberman, V.; Kang, M.; Ríos, C.; Richardson, K. A.; Rivero-Baleine, C.; Gu, T.; Zhang, H.; Hu, J., Electrically reconfigurable non-volatile metasurface using low-loss optical phase-change material. *Nat. Nano.* **2021**.
31. Wuttig, M.; Yamada, N., Phase-change materials for rewriteable data storage. *Nat. Mater.* **2007,** *6* (11), 824-832.
32. Lee, B.-S.; Abelson, J. R.; Bishop, S. G.; Kang, D.-H.; Cheong, B.-k.; Kim, K.-B., Investigation of the optical and electronic properties of Ge2Sb2Te5 phase change material in its amorphous, cubic, and hexagonal phases. *Journal of Applied Physics* **2005,** *97* (9), 093509.
33. Dong, W.; Liu, H.; Behera, J. K.; Lu, L.; Ng, R. J. H.; Sreekanth, K. V.; Zhou, X.; Yang, J. K. W.; Simpson, R. E., Wide Bandgap Phase Change Material Tuned Visible Photonics. *Advanced Functional Materials* **2019,** *29* (6), 1806181.
34. Delaney, M.; Zeimpekis, I.; Lawson, D.; Hewak, D. W.; Muskens, O. L., A New Family of Ultralow Loss Reversible Phase-Change Materials for Photonic Integrated Circuits: Sb2S3 and Sb2Se3. *Advanced Functional Materials* **2020,** *30* (36), 2002447.
35. Liu, H.; Dong, W.; Wang, H.; Lu, L.; Ruan, Q.; Tan, Y. S.; Simpson, R. E.; Yang, J. K. W., Rewritable color nanoprints in antimony trisulfide films. *Sci. Adv.* **2020,** *6* (51), eabb7171.
36. Hemmatyar, O.; Abdollahramezani, S.; Lepeshov, S.; Krasnok, A.; Brown, T.; Alu, A.; Adibi, A., Advanced Phase-Change Materials for Enhanced Meta-Display. *arXiv* **2021**, arXiv:2105.01313.
37. Daqiqeh Rezaei, S.; Dong, Z.; You En Chan, J.; Trisno, J.; Ng, R. J. H.; Ruan, Q.; Qiu, C.-W.; Mortensen, N. A.; Yang, J. K. W., Nanophotonic Structural Colors. *ACS Photon.* **2021,** *8* (1), 18-33.
38. Ren, H.; Fang, X.; Jang, J.; Bürger, J.; Rho, J.; Maier, S. A., Complex-amplitude metasurface-based orbital angular momentum holography in momentum space. *Nat. Nano.* **2020,** *15* (11), 948-955.
39. Dong, Z.; Bosman, M.; Zhu, D.; Goh, X. M.; Yang, J. K. W., Fabrication of suspended metal–dielectric–metal plasmonic nanostructures. *Nanotechnology* **2014,** *25* (13), 135303.
40. Dong, Z.; Wang, T.; Chi, X.; Ho, J.; Tserkezis, C.; Yap, S. L. K.; Rusydi, A.; Tjiptoharsono, F.; Thian, D.; Mortensen, N. A.; Yang, J. K. W., Ultraviolet Interband Plasmonics With Si Nanostructures. *Nano Lett.* **2019,** *19* (11), 8040-8048.
41. Jalali, M.; Yu, Y.; Xu, K.; Ng, R. J. H.; Dong, Z.; Wang, L.; Safari Dinachali, S.; Hong, M.; Yang, J. K. W., Stacking of colors in exfoliable plasmonic superlattices. *Nanoscale* **2016,** *8* (42), 18228-18234.
42. Palik, E. D., *Handbook of optical constants of solids* Academic Press: San Diego, 1998.
43. Philipp, H. R., Optical Properties of Silicon Nitride. *Journal of The Electrochemical Society* **1973,** *120* (2), 295.

*Supplementary Information for*

# Tunable Mie Resonances in the Visible Spectrum

Li Lu[1,#], Zhaogang Dong[2,#,*], Febiana Tijiptoharsono[2], Ray Jia Hong Ng[3], Hongtao Wang[1], Soroosh Daqiqeh Rezaei[1], Yunzheng Wang[1], Hai Sheng Leong[2], Joel K. W. Yang[1,2,*] and Robert E. Simpson[1,*]

[1]Singapore University of Technology and Design, 8 Somapah Road, 487372, Singapore

[2]Institute of Materials Research and Engineering, A*STAR (Agency for Science, Technology and Research), 2 Fusionopolis Way, #08-03 Innovis, 138634 Singapore

[3]Institute of High Performance Computing, A*STAR (Agency for Science, Technology and Research), 1 Fusionopolis Way, #16-16 Connexis, 138632 Singapore

[#]These authors equally contribute to this work.

*Correspondence and requests for materials should be addressed to R.E.S. (email: robert_simpson@sutd.edu.sg), J.K.W.Y. (email: joel_yang@sutd.edu.sg; telephone: +65 64994767) and Z.D. (email: dongz@imre.a-star.edu.sg ).

## S1. Reflection spectra for flat substrate

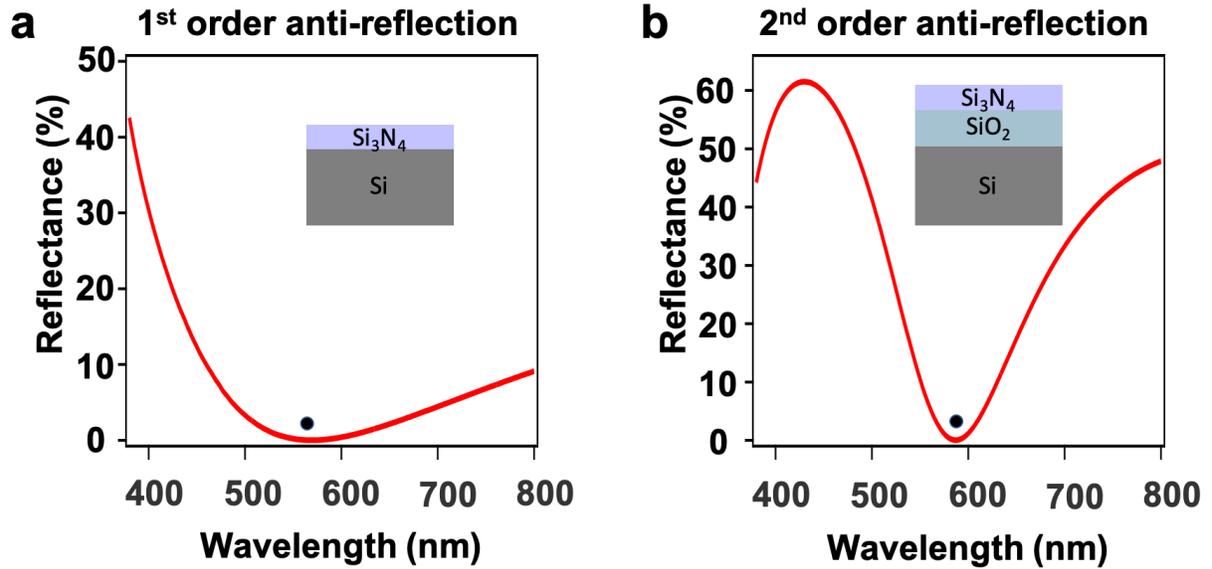

Fig. S1. Reflection spectra for flat substrate (a) 70 nm-$Si_3N_4$ coated Silicon substrate (1st order anti-reflection) and (b) 70 nm-$Si_3N_4$ coated 200 nm-$SiO_2$ on Silicon substrate (2nd order anti-reflection). The perfect anti-reflection wavelengths are highlighted by the black dots.

## S2. Nanofabrication process of Sb₂S₃ nanoantenna array.

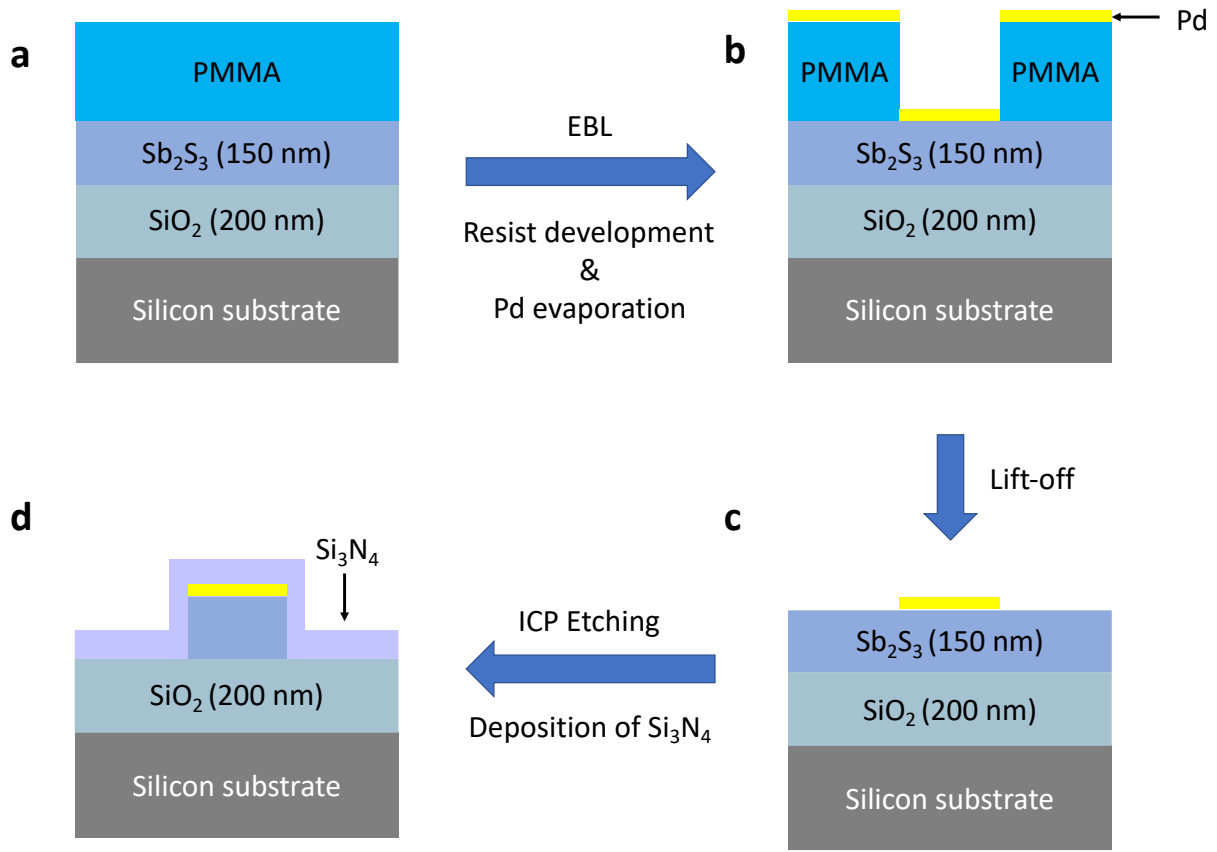

Fig. S2. Detailed schematic for illustrating the nanofabrication processes. (a) PMMA resist as spin coated onto silicon substrate with 200-nm-thick $SiO_2$ and 150-nm-thick $Sb_2S_3$ layer. (b) Electron beam lithography, resist development and Pd deposition using E-beam evaporator. (c) Lift-off process using NLP to make Pd hard mask. (d) Inductively-coupled plasma (ICP) etching with $Cl_2$ gas followed by $Si_3N_4$ deposition using ICP chemical vapor decomposition (CVD).

**S3. Sb₂S₃ nanoantenna color palette for amorphous and crystalline state in CIE chromaticity diagram.**

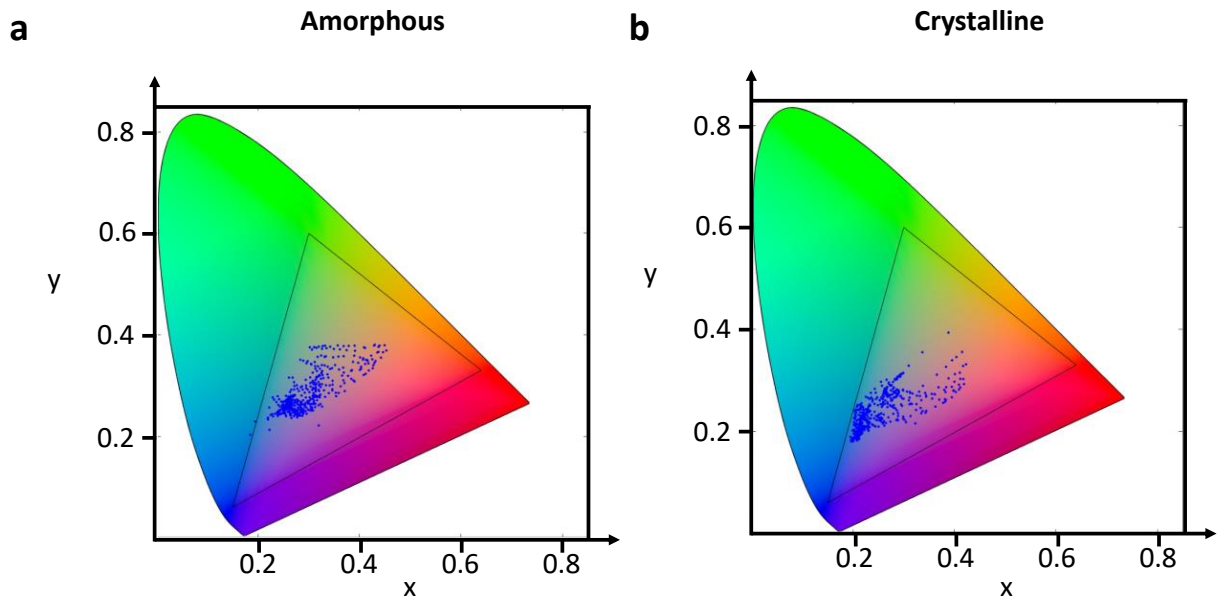

Fig. S3. Sb$_2$S$_3$ nanoantenna color palette in CIE 1931 chromatic diagram for (a) amorphous state, (b) crystalline state.

**S4. Simulated reflection spectra for Sb2S3 nanoantenna with a gap 90 nm and diameter 190 nm at both amorphous and crystalline states.**

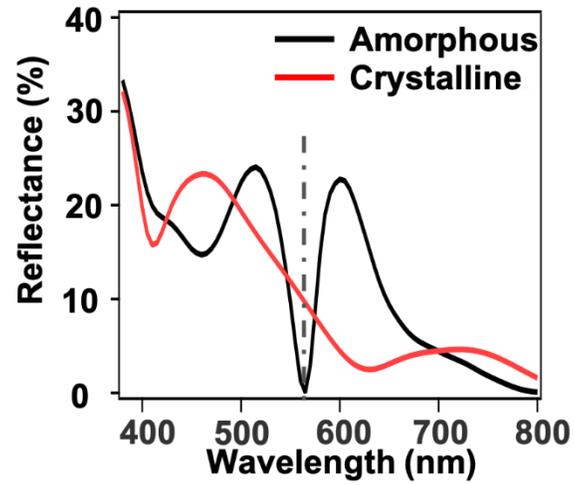

Fig. S4. Simulated reflection spectra for $Sb_2S_3$ nanoantenna with D = 190 nm, gap = 90 nm in the amorphous and crystalline phase.

**S5. Excited multipoles magnitude and corresponding phase in amorphous and crystalline phase.**

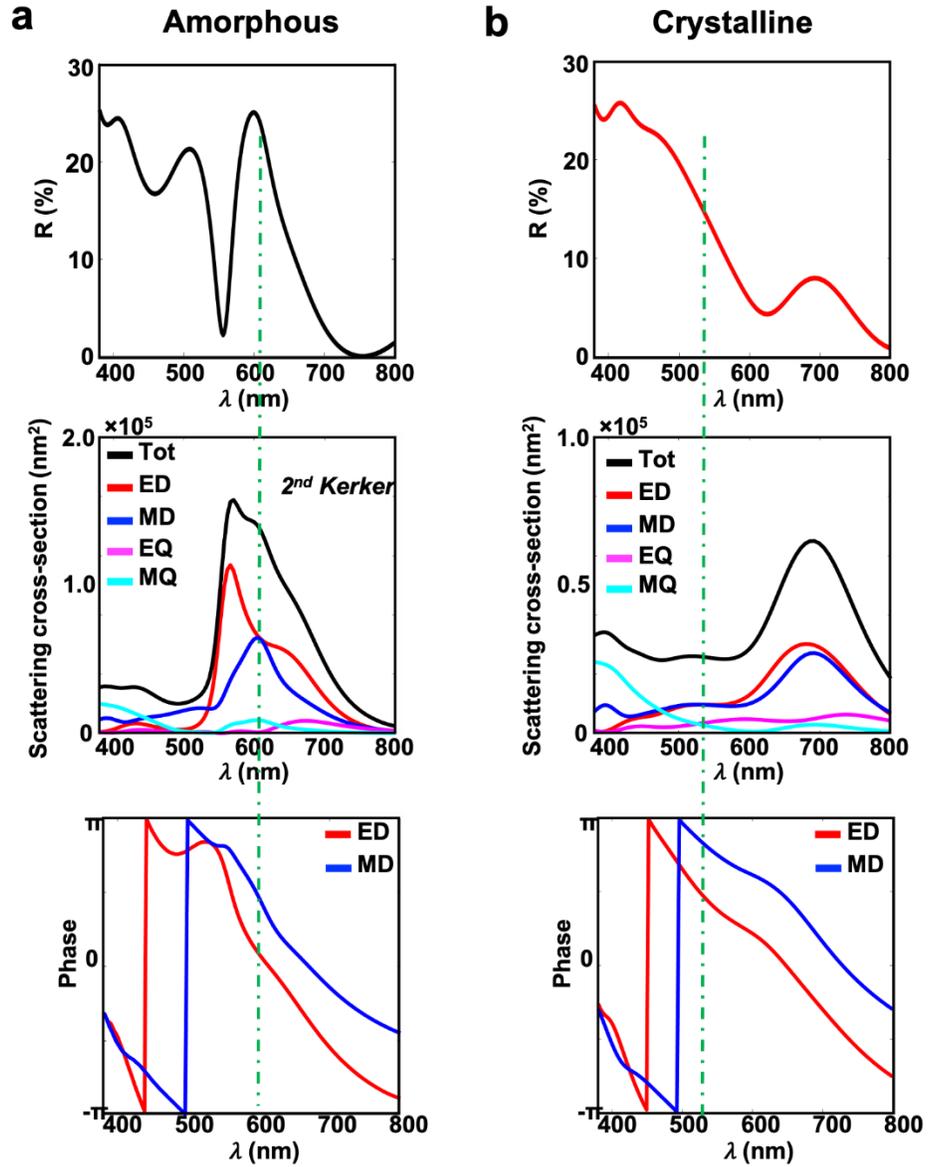

Fig. S5. Excited multipoles magnitude and corresponding phase in for $Sb_2S_3$ nanoantenna with D = 190 nm, gap = 90 nm in the amorphous and crystalline phase.

## S6. Sb₂S₃ nanoantenna color palette and SEM image before Si₃N₄ deposition.

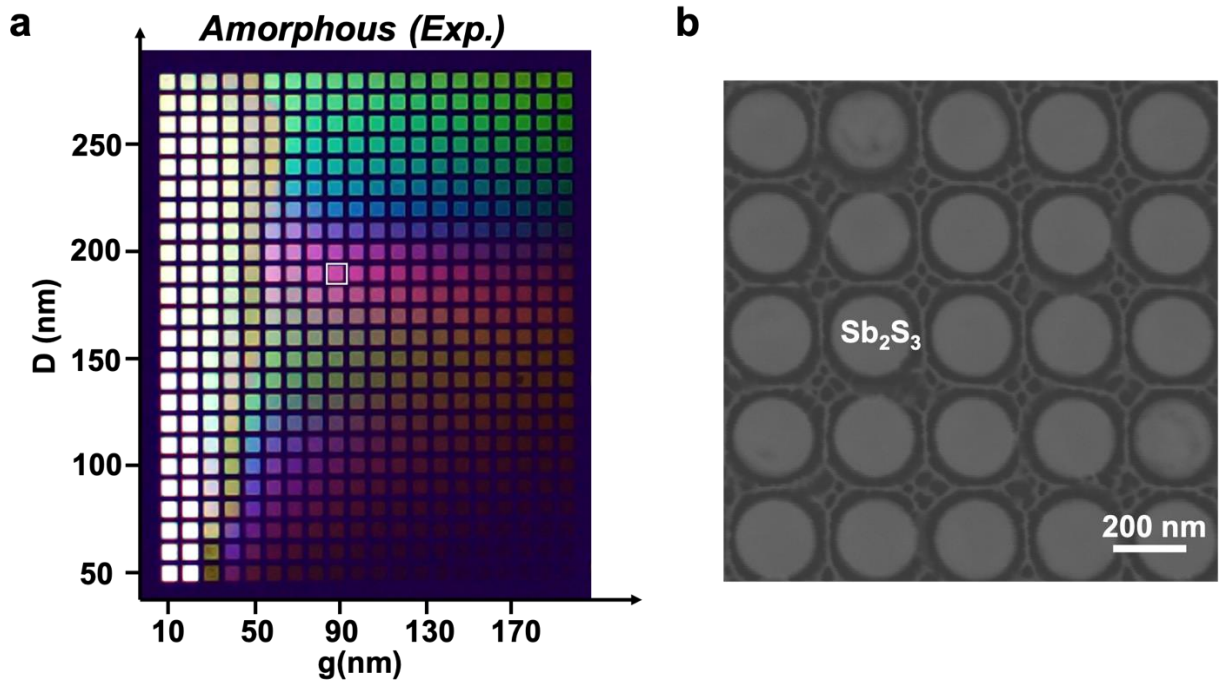

Fig. S6. (a) Bright field optical microscope images of the nanostructured color palette before depositing $Si_3N_4$ film. (b) SEM image for $Sb_2S_3$ nanoantenna with D = 190 nm, gap = 90 nm before $Si_3N_4$ deposition.

**S7. Simulated Sb₂S₃ color palette and reflection spectra for Sb₂S₃ nanoantenna with gap 90 nm, diameter from 70 nm to 280 nm.**

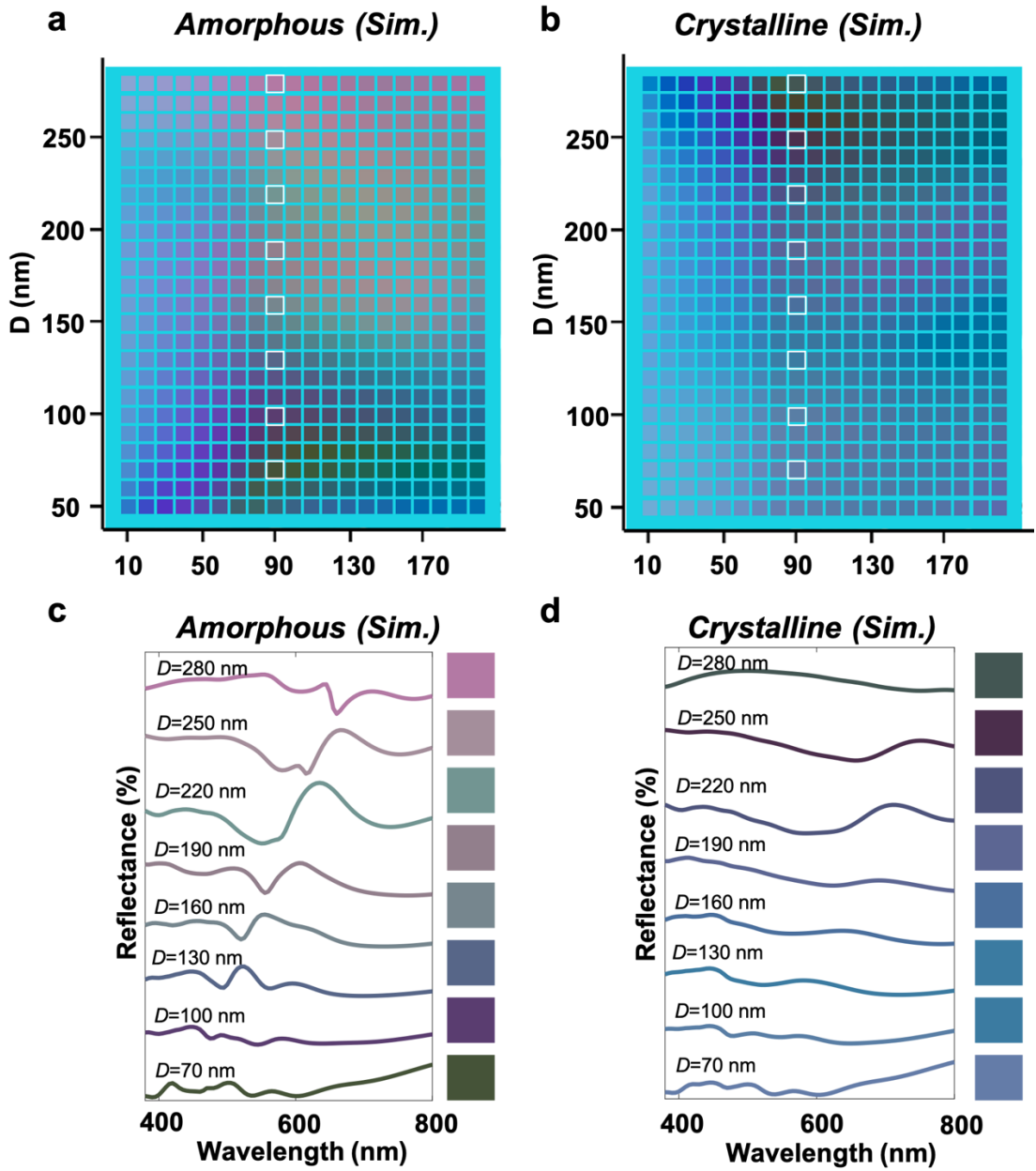

Fig. S7. (a-b) Simulated Sb$_2$S$_3$ color palette for amorphous and crystalline states, (c-d) Simulated reflection spectra for nanoantenna with gap = 90 nm, diameter from 70 nm to 280 nm for amorphous and crystalline states.

**S8. Scattering cross-section of Sb₂S₃ nanoantenna arrays in the amorphous and crystalline states.**

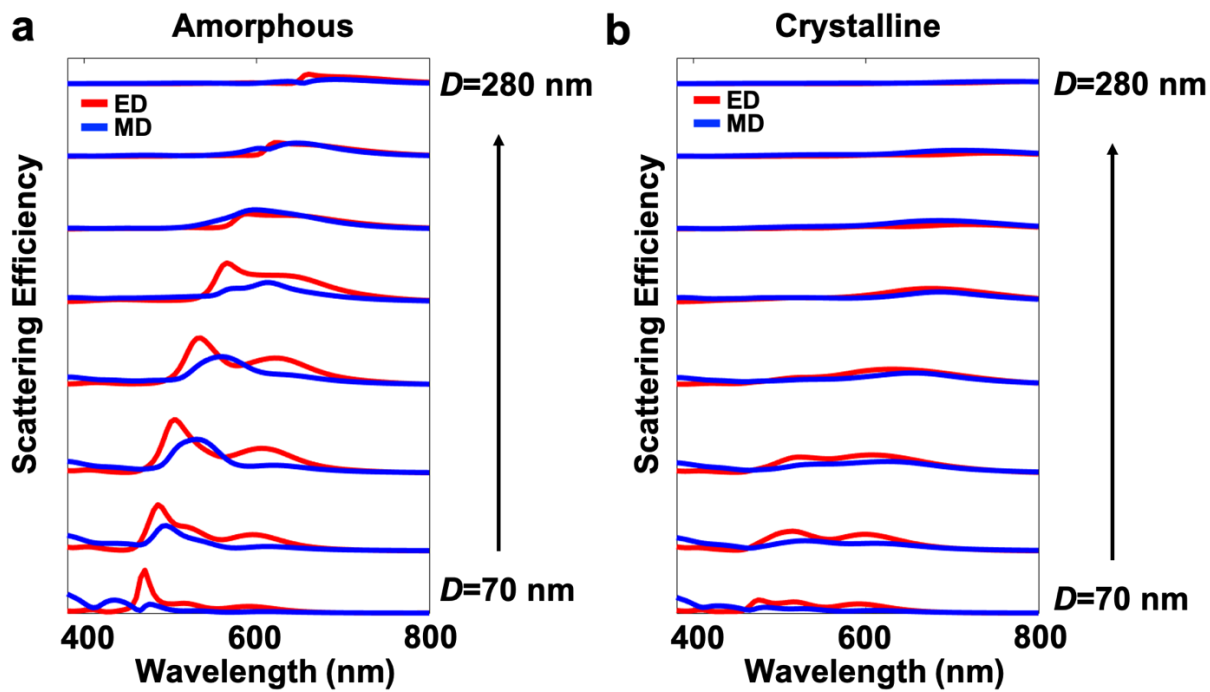

Fig. S8. Scattering cross-section of $Sb_2S_3$ nanoantenna arrays with, gap = 90 nm, diameter varied from 70 nm to 280 nm in the (a) amorphous and (b) crystalline phase.

## S9. Scattering cross-section of Sb₂S₃ nanoantenna single element in the amorphous and crystalline states.

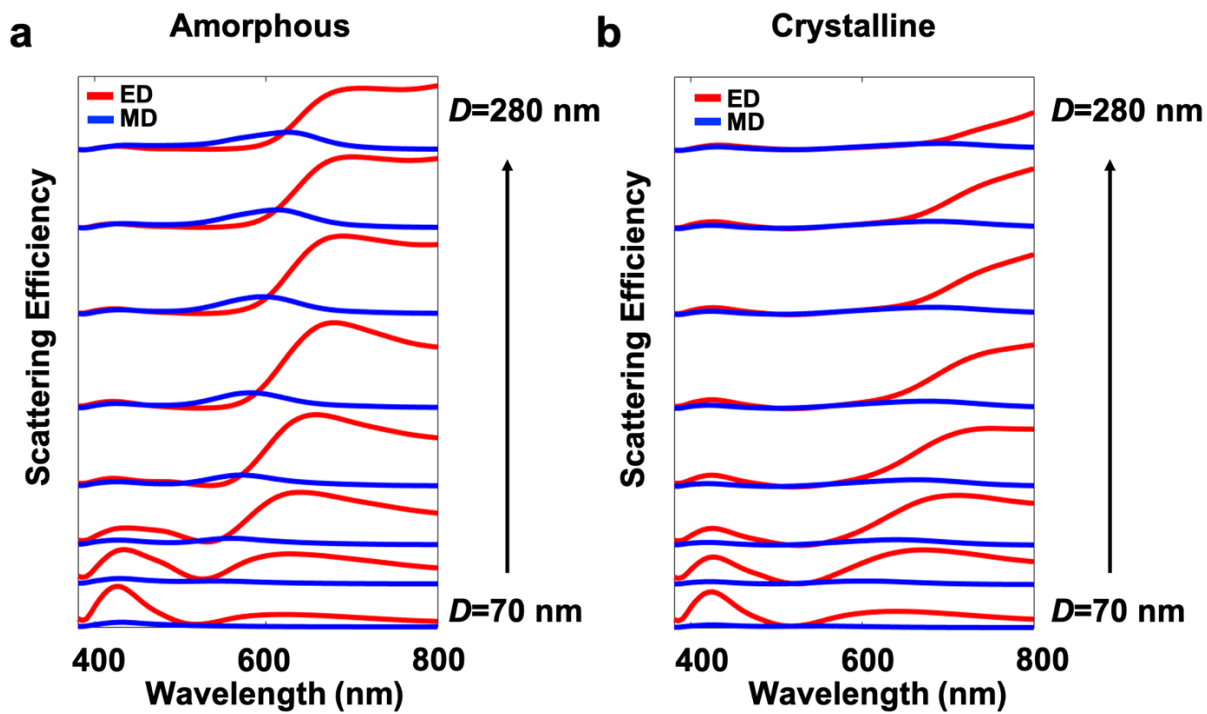

Fig. S9. Scattering cross-section of $Sb_2S_3$ nanoantenna single element with diameter varied from 70 nm to 280 nm in the (a) amorphous and (b) crystalline phase.